\documentstyle[12pt]{article}

\newcommand{\la}{\mbox{$\lambda$}}

\newcommand{\ra}{\mbox{$\rightarrow $}}

\newcommand{\pa }{\partial }

\newcommand{\al}{\alpha }

\newcommand{\ga}{\gamma }

\newtheorem{prop}{Proposition}
\newtheorem{thm}{Theorem}

\title{Trapped modes for periodic structures in waveguides}

\author{Julian Edward\footnotemark[2]}

\begin{document}
\maketitle
\begin{abstract}
The Laplace operator is considered for waveguides perturbed by a periodic 
structure consisting of $N$ congruent obstacles
spanning the waveguide.
Neumann boundary
conditions are imposed on the periodic structure, and either Neumann
or Dirichlet conditions on the guide walls. 
It is proven 
that there are at least $N$ (resp. $N-1$) trapped modes in the 
Neumann case (resp. Dirichlet case) under fairly general hypotheses,
including the special case where the obstacles consist of line segments
placed parallel to the waveguide walls.
\end{abstract}

\begin{section}{Introduction}
Let $N$ be a positive integer. Consider the region in ${\bf R}^2$:
\begin{equation}
\Omega =\{ (x,y): x\in (-\infty ,\infty ),\ y\in (0,2N)\} -\cup_{m=1}^{N}
{\cal O}_m,\label{region}
\end{equation}
with 
\begin{equation}
{\cal O}_m=\{ (x,y),\ x\in [-a,a],\ y\in 
[{2m-1}-g(x),{2m+1}+g(x)]\} .\label{ob}
\end{equation}
Here $a>0$, and $g$ is a  continuous function with $g(x)\in [0,1)$ and
 $g(\pm a)=0$.
Thus $\Omega$ can be viewed as a waveguide with $N$ congruent obstacles
placed periodically along the cross section. 
In \cite{LM}, Linton and McIvor studied the trapped modes in such regions
under the hypothesis that $g(x)$ was not identically zero.
Assuming Neumann boundary
conditions on the obstacles, and either Dirichlet or Neumann boundary
conditions on the guide walls, they proved the existence of at least
$N$ trapped modes in the Neumann case, and at least $N-1$ trapped modes
in the Dirichlet case. The existence of these trapped modes
was indicated earlier in numerical studies by 
Utsumomiya and Eatock Taylor \cite{UT} and Evans and Porter \cite{EP}.
These studies were motivated by a variety of  possible applications
to wave propagation in fluid and vibrating membranes; the reader is
referred to \cite{LM} for a thorough discussion of these.

Although the methods of Linton-McIvor apply for a wide variety on
assumptions of the geometry of the structure, they do not apply
to the important special case where the structure consists
of $N$ identical line segments placed parallel to the guide walls,
ie. $g\equiv 0$.

Furthermore, the numerical results in
 \cite{UT}, \cite{EP} also fail to indicate any trapped modes in
this setting.
The main purpose 
of this note is prove the existence of at least $N$ (resp. $N-1$)
trapped modes in this setting for the Neumann (resp. Dirichlet) case.
The methods of this paper also
apply to the more general periodic structures described by Eqs.~\ref{region},
~\ref{ob},
and this work might also be
of interest because the upper
bounds proven here on the associated frequencies will sometimes be
sharper than those found  in Linton-McIvor.

In addition, we consider the trapped modes of the region 
\begin{equation}
\tilde{\Omega }
=\{ (x,y): x\in (-\infty ,\infty ),\ y\in (0,2N)\} -\cup_{m=1}^{N-1}
{\cal O}_m,\label{region2}
\end{equation}
with 
\begin{equation}
{\cal O}_m=\{ (x,y),\ x\in [-a,a],\ y=2m\} .\label{obs2}
\end{equation}

For such regions with Neumann boundary conditions both on the 
obstacles and the guide walls, we prove the existence of 
$N-1$ embedded eigenvalues. The methods of this paper seem to
fail in this case when one has Dirichlet boundary conditions on
the  guide walls.

We note that a  number of papers have proven the existence of 
at least $one$ trapped mode for the case of 
a $single$ line segment is placed parallel to the guide walls:
see \cite{DP},\cite{U}, \cite{V}, \cite{E}, \cite{P},
 and the references found therein.

To prove the announced results,
we use the symmetry of the problem to 
decompose the ambient Hilbert space into a direct sum of $N+1$ invariant
subspaces, as in \cite{LM}. 
However, we differ in our choice of test functions.
In the direction transversal to the guide walls, Linton and McIvor's
test function is essentially sinusoidal. In this paper, we choose
a test function which has a jump discontinuity across the obstacles, 
and which  arises naturally from the Hilbert space decomposition.

\end{section}
\begin{section}{Statement of results and proofs}

We define the Laplace operator as
$$\Delta =-\frac{\pa^2}{\pa x^2}-\frac{\pa^2}{\pa y^2}.$$
Let $C_b^{\infty}(\Omega )$ be the smooth functions of bounded support
on $\Omega$.
We shall study the self-adjoint operators $\Delta_1$, $\Delta_2$
on $L^2(\Omega )$, where $\Delta_1$ has operator core 
$$\{ u\in C_b^{\infty}(\Omega ): \frac{\pa u}{\pa \eta }|_{\pa \Omega}=0\} ,$$
and $\Delta_2$ has operator core
$$\{ u\in C_b^{\infty}(\Omega ): \frac{\pa u}{\pa \eta }|_{\pa {\cal O}_j}=0,
j=1,\ldots ,m,\ u|_{y=0}=u|_{y=2N}=0 \} .$$
Of course, $\Delta_1$ is simply the Neumann Laplacian, while we shall
refer to $\Delta_2$ as the Dirichlet case.

\begin{subsection}{Neumann boundary conditions}
In this section we prove the following:
\begin{thm}
Suppose $\Omega$ satisfies Eqs.~\ref{region},~\ref{ob}. Then,
the operator $\Delta_1$ has at least  $N$ linearly independant trapped 
modes. Among the trapped modes, there exist $N$ whose associated 
frequencies, denoted $\mu_m$, satisfy
$$\mu_m\leq (\frac{m\pi}{2N})^2,\ m=1,\ldots, N.$$
\end{thm}
We begin the proof of the theorem by recalling the decomposition,
used in \cite{LM}, of $L^2(\Omega )$ 
into $\Delta_1$-invariant 
subspaces.
Suppose $f\in L^2[0,2N]$. We extend $f$ to $L^2(-\infty, \infty )$ using the rules
$$f(y)=f(-y),\ f(2N-y)=f(2N+y).$$
Set
$f(y)=\sum _{m=0}^Nf_n(y)$, where
\begin{equation}
f_m(y)= \frac{\ga_m^N}{2N}\sum_{n=-N}^{N-1}c_{mn}^Nf(y+2n),\label{fm}
\end{equation}
with 
\begin{equation}
c_{m,n}^N=\cos (mn\pi /N),\label{trig}
\end{equation}
 and 
$\gamma_m^N=2/(1+\delta_{m0}+\delta_{mN})$.
Here $\delta_{ij}$ is the Kronecker delta function.
It is shown in \cite{LM} that $f(y)=\sum_{m=0}^Nf_m(y)$, and  in
fact 
$$L^2(\Omega )=S_0\oplus \ldots S_N,$$
with the $S_m$ the image of the mapping $f\ra f_m$. The subspaces
$\{ S_m\} $ are mutually orthogonal and are invariant under the Laplacian.
We label 
$\Delta_1 |_{S_m}\equiv A_m$. Then
$$\inf \sigma_{ess}(A_m)=\frac{m^2\pi^2}{4N^2}.$$ 

Recall the Rayleigh quotient is given by 
\begin{equation}
Q(\phi )=\frac{\int_{\Omega}|\nabla \phi|^2}{\int_{\Omega }|\phi |^2},
\phi \neq 0,\label{ray}
\end{equation}
where $\phi $ is in the quadratic form domain of  $A_m$. For Neumann
boundary conditions, the quadratic form domain of $A_m$ is 
$$S_m\cap H^1(\Omega )=\{ u\in S_m: |\nabla u|\in L^2(\Omega )\} .$$

To prove the existence of eigenvalues below the essential spectrum of 
$A_m$, (and hence the existence of trapped modes for $\Delta_1$),
it suffices to construct
$\phi$ such that $Q(\phi )< \frac{m^2\pi^2}{4N^2}.$

Fix $x$ 
and $m$, and for notational simplicity  set $c_{mn}^N=
c_n$. We label the 
intervals $(0,1-g(x)),\ (1+g(x),3-g(x)),\ldots ,(2N-1+g(x),2N)$ 
as $I_0,I_1,\ldots,I_{N}$ respectively. 

Let $\tilde{v}(y)=1$ on $I_0$ and 0 elsewhere.
Let $v$ be the image of 
$\tilde{v}$ under the 
 mapping $f\ra f_m$. Then, by  Eq.~\ref{fm}, 
\begin{equation}
v(y)=\frac{\ga_m^N}{2N}\cos(\frac{m\pi}{N}\cdot j)\ \mbox{for }y\in 
I_j.\label{v1}
\end{equation}

Let $b\in [0,a)$, let $\al >0$, and 
define a piecewise differentiable functions $\chi$ and $\psi_{\al}$ on $\Omega $ by 
$$\chi (x)=\left \{  
\begin{array}{cc}
0, & |x|>a,\\
1, & |x|<b,\\
\frac{a-x}{a-b}, & x\in (b,a),\\
\frac{a+x}{a-b}, & x\in (-a,-b);
\end{array}
\right . 
$$
$$\psi_{\al} (x)=\left \{  
\begin{array}{cc}
e^{-\al (|x|-a)}, & |x|>a,\\
1, & |x|\leq a.\\
\end{array}
\right .
$$

Our test function for Eq.~\ref{ray} will be:
$$\phi (x,y) =\chi (x) v(y)+\la \psi_{\al}(x)\cos (\frac{m\pi y}{2N}),$$
where $\la >0$ is a parameter to be chosen later.
Since $v$ is in the image of the mapping $f\ra f_m$, it follows that
$\chi v \in S_m$. Also, 
$\psi_{\al}(x)\cos (\frac{m\pi y}{2N})\in S_m$ by \cite{LM}, Eqs.2.12, 2.13.
Thus $\phi$ is in the quadratic form domain of $A_m$.
This test function can be compared to the one used 
in \cite{LM}, Eq.4.13.

In what follows, it is convenient to set 
$$\| {\bf v}\|_2^2
\equiv
|v_0|^2+2|v_2|^2+2|v_3|^2+\ldots + 2|v_{N-1}|^2+|v_N|^2,$$
where $v_j\equiv v|_{I_j}$.
\begin{prop}
Let $p=\frac{m\pi}{2N}$. Then:
$$
\frac{\int |\nabla \phi |^2}{\int |\phi |^2}-p^2=
\frac{\la^2\al N+
\int_{x=-a}^a\left ( \| {\bf v}\|_2^2
(1-g)((\chi')^2-p^2\chi^2)-\la
\frac{CN^2p^2}{m\pi}\chi \sin (p(1-g))\right )dx}
{\la^2N/\al + 
\int_{x=-a}^a\left (\la^2N(1-g(x))+\frac{C\la N^2}{m\pi}
\sin (\frac{m\pi}{2N}(1-g))\chi
+\| {\bf v}\|_2^2\chi^2 (1-g(x))\right ) dx};
$$
here $C=4$ for $m=1,\ldots ,N-1$ and $C=8$ for $m=N$.
\end{prop}
The proof the this result appears in the appendix.

We now complete the proof of the theorem.
By Eq.~\ref{ray} and the remarks that follow it,
it suffices for the right hand side of the last equation to be negative.
Note that the denominator is positive, and the same holds for the
term $\int_{|x|<a}\chi (x) \sin (p(1-g(x)))dx$. Choose $\la$ sufficiently large
that 
$$ \|{\bf v}\|_2^2\int_{|x|<a}(1-g)((\chi')^2-p^2\chi^2)-\la
\frac{CN^2p^2}{m\pi}\int_{|x|<a}\chi \sin (p(1-g))<0.$$
Fixing this $\la$, we then choose $\al>0$ so that
$$
\| {\bf v}\|_2^2\int_{|x|<a}(1-g)((\chi')^2-p^2\chi^2)-\la
\frac{CN^2p^2}{m\pi}\int_{|x|<a}\chi \sin (p(1-g))+\la^2\al N <0.$$
The theorem is proven.

\end{subsection}
\begin{subsection}{Dirichlet case}
In this section we prove the following:
\begin{thm}
Suppose $\Omega$ satisfies Eqs.~\ref{region},~\ref{ob}. Then,
the operator $\Delta_2$ has at least  $N-1$ linearly independant trapped 
modes. Among the trapped modes, there exist $N-1$ whose associated 
frequencies $\mu_j$ satisfy
$$\mu_m\leq (\frac{m\pi}{2N})^2,\ m=1,\ldots, N-1.$$
\end{thm}
For
 Dirichlet boundary conditions, 
we extend $f\in L^2(0,2N)$ to $L^2(-\infty ,\infty)$ via the equations
$$f(-y)=-f(y),\ f(2N+y)=-f(2N-y).$$
Then, using Eq.~\ref{fm} as in the Neumann case, we have the decomposition
$L^2(\Omega )=\oplus_{m=0}^NS_m$, where the $S_m$ are mutually orthogonal
and $S_m$ are invariant under $\Delta_2$. Setting
$A_m\equiv \Delta_2|_{S_m}$, we have
$$
\inf \sigma_{ess}(A_m)=
\left \{ 
\begin{array}{cc}
\pi^2, & m=0\\
(\frac{m\pi}{2N})^2, & m=1,\ldots ,N.
\end{array}
\right .
$$
As in the Neumann case, we construct a test function for the Rayleigh 
quotient. In this case the quadratic form domain for $A_m$ is 
the closure in $H^1(\Omega )$-norm of the set
$$\{ u\in S_m\cap C_b^{\infty}(\Omega ): \mbox{support}(u)\cap \{ y=0\}
=\mbox{support}(u)\cap \{ y=2N\}=\phi \}.$$

Let $\{ I_j\}$ be as in the Neumann case.
Let 
$$\tilde{v}_j (y)=\left \{  
\begin{array}{cc}
1, & y\in I_j  ,\\
0, & \mbox{elsewhere.}
\end{array}
\right . 
$$
Denote by $v_j$ the image of $\tilde{v}_j$
under the 
 mapping $f\ra f_m$. Then,  for $j=1,\dots ,N-1$, 
we have by Eq.~\ref{fm},
\begin{eqnarray}
v_j(y)& =& \frac{\ga_m^N}{2N}(c_{j-s}-c_{-j-s}),\mbox{ for }
y\in I_s\nonumber \\
&  =&  \frac{\ga_m^N}{N}\sin(\frac{m\pi}{N}\cdot j)
\sin (\frac{m\pi}{N}\cdot s)\mbox{ for }y\in I_s.\label{v2}
\end{eqnarray}
Here we have used
$c_J=c_{-J}$,  
$c_{J-2N}=c_J$, and the identity $\cos (A-B)-\cos (A+B)=2\sin A\sin B.$

We chose $j$ so that $\sin(\frac{m\pi}{N}\cdot j)\neq 0$. 
Thus $v_j\equiv 0$ if and only if $m=0$ or $m=N$. 

The test function is defined as:
$$\phi (x,y) =\chi (x) v_j(y)+\la \psi_{\al}(x)\sin (\frac{m\pi y}{2N}),$$
where $\la >0$ is a parameter to be chosen later, and $\chi ,\psi_{\al}$
as in the Neumann case. 
Note that $\chi v_j \in S_m$, and also that $\chi v_j$ vanishes at $y=0$,
$y=2N$ (see Eq.~\ref{v2}). Hence, $\chi v_j$ is in the quadratic form domain
of $A_m$. Also, by (\cite{LM}, Eq.2.15),
 $\psi_{\al}(x)\sin (\frac{m\pi y}{2N})$
is in the quadratic form domain of $A_m$, and hence 
$\phi$ is in
the quadratic form domain of $A_m$.
 
The proof of the theorem now follows from a 
word to word repetition of the argument used in the
Neumann case.

\end{subsection}
\begin{subsection}{Line segments placed on $y=2,4,\ldots , 2N-2$}
In this section, we prove
\begin{thm}
Suppose $\Omega$ satisfies Eqs.~\ref{region2},~\ref{obs2}. Then,
the operator $\Delta_1$ has at least  $N-1$ linearly independant trapped 
modes. Among the trapped modes, there exist $N-1$ whose associated 
frequencies $\mu_m$ satisfy
$$\mu_m\leq (\frac{m\pi}{2N})^2,\ m=1,\ldots, N-1.$$
\end{thm}
First, we note that the region $\Omega$ under these hypotheses
satisfies the conditions necessary for the decomposition
$$L^2(\Omega )=S_0\oplus \ldots S_N,$$
with $f\ra f_m$ defined exactly as above (see \cite{LM}, p.3).

We label the intervals $(0,2),\ (2,4),\ldots ,(2N-2,2N)$ 
as $I_1,I_2,\ldots,
I_{N}$ respectively. 
Let $\tilde{v}=1$ on $I_1$, and $\tilde{v}=0$ elsewhere.

Denote by $v$ the image of $\tilde{v}$
under the 
 mapping $f\ra f_m$. Then
we have by Eq.~\ref{fm},
\begin{eqnarray*}
v(y)& =& \frac{\ga_m^N}{2N}(c_{-1-s}+c_{-s}),\mbox{ for }y\in I_s\\
&  =&  \frac{\ga_m^N}{N}\cos (\frac{m\pi}{N}\cdot \frac{1}{2})
\cos (\frac{m\pi}{N}\cdot (\frac{1}{2}+s)),\mbox{ for }y\in I_s.\label{v3}
\end{eqnarray*}
Here we have used $\cos (A-B)+\cos (A+B)=2\cos A\cos B$.

Note that 
$v\equiv 0$ if and only if $m=N$. 

The test function is defined as:
$$\phi (x,y) =\chi (x) v(y)+\la \psi_{\al}(x)\cos (\frac{m\pi y}{2N}),$$
where $\la >0$ is a parameter to be chosen later, and $\chi ,\psi_{\al}$
as in the Neumann case.
The proof of the theorem now follows by mimicking 
 the argument used in the
Neumann case.
 
\end{subsection}
\end{section}
\begin{section}{Appendix}

In this section we prove Proposition 1.

We use the notation of Section 2.1. In what follows, we denote
by $\{ |x|<a\}$ the set $\{ (x,y): |x|<a\}$.
We have
\begin{eqnarray*}
\int_{\{ |x|>a\} } | \phi |^2 
& =& 2\la^2\int_{x=a}^{\infty}\int_{y=0}^{2N}
\cos^2(\frac{m\pi y}{2N})
e^{-2\al (x-a)}\ dy \ dx\\
& = & \la^2N/\al .
\end{eqnarray*}
Next, we calculate:
\begin{eqnarray*}
\int_{\{ |x|<a\} }| \phi |^2 & = & 
\la^2\int_{\{ |x|<a\} }
\cos^2(\frac{m\pi y}{2N})
+\int_{\{ |x|<a\} }v^2\chi^2
+2\la \int_{\{ |x|<a\} }\chi v\cos (\frac{m\pi y}{2N}).\label{three}
\end{eqnarray*}
The first of the integrals on the right hand side of  
the last equation we compute as
follows:
\begin{eqnarray}
\int_{\{ |x|<a\} }
\cos^2(\frac{m\pi y}{2N})& =& \int_{-a}^a\left (
\int_{y=0}^{1-g(x)}\cos^2(\frac{m\pi y}{2N})\ dy \right . \nonumber \\
& & \left . +
\sum_{n=1}^{N-1}\int_{y=2n-1+g(x)}^{2n+1-g(x)}\cos^2(\frac{m\pi y}{2N})\ dy
+
\int_{2N-1+g(x)}^{2N}\cos^2(\frac{m\pi y}{2N})\ dy
\right ) \ dx
 \nonumber \\
 & = & N\int_{x=-a}^a
\left ( (1-g(x))+\frac{\sin(\frac{m\pi (1-g)}{N})}{m\pi }\cdot 
(\sum_{i=0}^{N-1}\cos(\frac{m2\pi i}{N})) \right )\ dx\nonumber \\
  & = & N\int_{x=-a}^a(1-g(x))dx.\label{cos}
\end{eqnarray}
Here we have used the identity
$\sum_{i=0}^{N-1}\cos(\frac{m2\pi i}{N})=0$, which follows from
$\sum_{j=0}^{N-1}e^{2ijm\pi /N}=0$.

For the third integral on the right hand side of Eq.~\ref{three}, a 
similar computation yields
\begin{eqnarray*}
\int_{\{ |x|<a\} }
\chi v\cos (\frac{m\pi y}{2N})
& = & \int_{ x=-a }^a\frac{2N}{m\pi}\chi \sin (\frac{m\pi}{2N}(1-g))(v_0+
2\sum_{i=1}^{N-1}v_i\cos (\frac{2m\pi i}{2N})+v_N)\ dx\\
& = &\int_{ x=-a }^a\frac{2N^2}{m\pi}\chi \sin (\frac{m\pi}{2N}(1-g))\  dx\ \mbox{ if }m=1,\ldots , (N-1),\\
& = &
\int_{x=-a }^a
\frac{4N^2}{m\pi}\chi \sin (\frac{m\pi}{2N}(1-g))\ dx \ \mbox{ if }m=0,N.
\end{eqnarray*}
Also, a straightforward calculation yields that
$$\int_{\{ |x|<a\} }
v^2\chi^2 =\| {\bf v}\|_2^2\int_{-a}^a \chi^2 (1-g(x))dx.$$
Thus,
\begin{equation}
\int_{\Omega}|\phi |^2
=\la^2
N/\al+\int_{x=-a}^a\left ( \la^2N(1-g(x))+\frac{C\la N^2}{m\pi}
\sin (\frac{m\pi}{2N}(1-g))
+\| {\bf v}\|_2^2\chi^2 (1-g(x))\right ) dx.\label{num}
\end{equation}
Here, $C=8$ for $m=0,N$ and $C=4$ otherwise.

In what follows,it is convenient to set $\frac{m\pi}{2N}=p$. Then
\begin{eqnarray}
\int_{\Omega}|\nabla \phi |^2 & = 
& \int_{\Omega}\left ( \la^2p^2\sin^2 (py)\psi_{\al}^2+\la^2\cos^2 (py)(\psi_{\al}')^2
\right .\nonumber \\
& &\left . +v^2(\chi')^2+2\la v\cos (py)\psi_{\al}'\chi'\right ).\label{big}
\end{eqnarray}
The fourth integrand is identically zero because $\psi_{\al}'
\chi '=0$.
Next, we calculate the first three terms in Eq.~\ref{big}.
First,
\begin{eqnarray*}
\la^2p^2\int_{\Omega}\sin^2(py)\psi_{\al}^2& = &
\la^2p^2\int_{\{ |x|<a\} }
(1-\cos^2(py))+2\la^2p^2\int_{x=a}^{\infty}\int_{y=0}^{2N}
e^{-2\al (x-a)}\sin^2(py) \ dy\ dx \\
& =& \la^2p^2N\int_{x=-a}^a(1-g(x))dx+\la^2p^2N/\al ;
\end{eqnarray*}
here we have used Eq.~\ref{cos}.
Similarly,
$$\la^2\int_{\Omega}\cos^2(py)(\psi_{\al}')^2=\la^2N\al ,$$
and
$$\int_{\Omega}v^2(\chi')^2=\| {\bf v}\|_2^2\int_{|x|<a}(\chi')^2(1-g(x))dx.$$
Thus 
\begin{equation}
\int_{\Omega}|\nabla \phi|^2=\la^2p^2N\int_{|x|<a}(1-g(x))dx+\la^2p^2N/\al 
+\la^2N\al+
\| {\bf v}\|_2^2\int_{|x|<a}(\chi')^2(1-g(x))dx.\label{denom}
\end{equation}
The proposition now follows from Eqs.~\ref{denom},~\ref{num}.
\end{section}

\end{document}